\documentclass[twoside]{elsart}

\usepackage{times}
 \usepackage{graphics}
\usepackage{epsfig}
\usepackage{amsfonts}
\usepackage{amssymb}

\renewcommand{\vec}[1]{\mbox{\boldmath$#1$}}

\newcommand{\eps}{\varepsilon}

\begin{document}

\begin{frontmatter}

\title{Simulation of Phase Combinations in Shape Memory \\
Alloys Patches by Hybrid Optimization Methods }

\author[A1]{Linxiang Wang} and
  \ead{wanglinxiang@mci.sdu.dk}
\author[A2]{Roderick V. N. Melnik}
\ead{rmelnik@wlu.ca}

\address[A1]{Mads Clausen Institute, Faculty of Engineering, \\
 University of Southern Denmark,\\
Sonderborg, DK-6400, Denmark }
\address[A2]{Mathematical Modelling and Computational Sciences
 \\ Wilfrid Laurier University, Waterloo, \\
75 University Ave W, Canada N2L 3C5 }


\begin{abstract} In this paper, phase combinations among martensitic variants
in shape memory alloys patches and bars are simulated by a hybrid optimization
 methodology. The mathematical model is based on the Landau theory of phase transformations. Each stable
phase is associated with a local minimum of the free energy function, and the phase
combinations are simulated by minimizing the bulk energy.  At low temperature, the free
energy function has double potential wells leading to non-convexity of the
optimization problem.  The methodology proposed in the present paper is based
on an initial estimate of the global solution by a genetic algorithm, followed
by a refined quasi-Newton procedure to
locally refine the optimum. By combining the local and global search algorithms, the
phase combinations are successfully simulated.  Numerical
experiments are presented for the phase combinations in a SMA patch under
several typical mechanical loadings.
\end{abstract}

\end{frontmatter}


\section{Introduction}

Shape Memory Alloys (SMA) are materials with increasing range of applications
 in engineering, aerosapce, and biomedical industries. They possess unique properties of
 being able to recover their
original shape after permanent deformations. These materials can directly transduce
thermal energy into mechanical and vice versa. The key to ``pseudo-elastic"
 behaviour of these materials and the ``shape
memory effect" \cite{Birman1997,Wayman1998} is held by
 the first-order martensitic phase transformations in these materials. Indeed,
drastic changes in their properties are originated from  their microstructure
  or phase combinations.  Via the phase transformation,  the microstructure of
the material can be switched among various combinations between austenite, martensite
variants, or  their mixtures. Austenitic phase is a more symmetric phase of the
 crystallic lattice, prevailing at high temperature, while martensite is a
 less symmetric, low-temperature phase
 \cite{Lookman2003b,Birman1997,Wayman1998}.  Between these two critical
 situations, austenite and martensite might co-exist providing a typical
 example of phase combinations.  Upon external loading, one phase combination can
be switched to another. If the material is constrained at the boundary,  a
 specific (``self-accommodating'') combination of different phases  will be
 established such that the bulk energy in the constrained
domain will be minimized \cite{Lookman2003b,Lookman2003,Luskin1996}.

The mathematical framework for modeling phase combinations in shape memory
materials  is based on the solution of the variational problem
 with respect to  a frame-indifferent
non-convex free energy function  $\phi(Y,\theta)$
(\cite{Ball1987,Ball1992,Collins1993,Luskin1996,Lookman2003} and references therein) :
\begin{equation}
 W(Y)=\int_{\Omega}\phi(Y,\theta)dV,
 \label{eq1-1}
\end{equation}
\noindent where $\Omega$ is the reference configuration associated with the
considered material, $Y$ is the deformation tensor, and $\theta$ is the
temperature of the material. Hence, by minimizing $W(Y)$ from (\ref{eq1-1}), we minimize the bulk energy of the considered
structure, as a functional of the deformation tensor,  at temperature
$\theta$.  This procedure, performed particularly often
at the mesoscale
\cite{Bartels2004,Carstensen2005,Collins1993,Luskin1996,Lookman2003}, has
several known difficulties. It is known that in the general case  the
variational problem given by Eq.(\ref{eq1-1}) may have infinitely many minimizers
(\cite{Bartels2004,Luskin1996,Carstensen2005} and references therein). On the
other hand, a more
precise definition of the free energy that allow us to account for interfacial energy
effects by introducing gradients of the order parameters is a highly
non-trivial task (\cite{Lookman2003b,Lookman2003} and references therein)
connected with additional difficulties. The problem can also be regularized by
assigning simplified, e.g. affine,  boundary conditions.  For example, for the simulation of
simple laminated microstructure, the affine boundary
conditions  are constructed by assuming that the deformation gradient of the material on
the domain boundary is a linear combination of its equilibrium deformation gradients:
\begin{equation}
Y(x)=(\lambda F_{0}+(1-\lambda)RF_{1})x,\, x\in\partial\Omega,
\label{eq1-2}
\end{equation}
\noindent  where $R$ is a rotation matrix  satisfying a twinning
equation  (see \cite{Bartels2004,Carstensen2005,Collins1993,Luskin1996}) and
$\lambda$ is the thickness of layers in the laminated microstructure.  $F_0$ and $F_1$ are
the deformation gradients  that minimize the local energy function  (for square to
rectangular transformation)  when the deformation
gradient $\nabla Y$ takes any of the following following values:
\begin{eqnarray}
F_{0}=\left(\begin{array}{cc}
1+e_{m}/\sqrt{2} & 0\\
0 & 1-e_{m}/\sqrt{2}\end{array}\right), &
F_{1}=\left(\begin{array}{cc} 1-e_{m}/\sqrt{2} & 0\\
0 & 1+e_{m}/\sqrt{2}\end{array}\right),
\label{eq1-3}
\end{eqnarray}
\noindent where $e_{m}$ is a local minimum of the  free energy function
$\phi(Y,\theta)$. In the general case, however, the above affine boundary conditions
may not be appropriate. 

In solving problem  (\ref{eq1-1}), we have to face also numerical
 challenges connected with several local minima, resulted from phase mixtures
 under low and moderate temperature regimes, and non-convexity of the
problem (\cite{Alfio2000,Carstensen2005,Luskin1996,Yuan1997} and references therein).
Minimization procedures based on conventional local
search methods with randomly chosen  initial guesses  \cite{Bhattacharya1999} may not lead to a satisfactory result
 in those cases where the solution space is discretized with a large number of
 node points.

In the present paper, we construct a mathematical  model for the simulation of
phase combinations in SMA materials on the basis of the Landau
theory. We associate the phase
combination with the global minimizer of the bulk energy
 of  the SMA structure with the prescribed mechanical boundary conditions.
We develop a hybrid optimization
strategy consisting of two main steps. Firstly, we apply the Genetic Algorithm
(GA) and its global exploration capacity to obtain an initial estimation of the
global minimizer.  Then, we apply the quasi-Newton method to refine
 such an estimation locally. Finally, the developed procedure is demonstrated
 by several numerical examples simulating phase
 combinations in SMA materials.


\section{Landau Free Energy Function and Variational Formulation}

\begin{figure}     \begin{center}
  \includegraphics[height=6cm,width = 10cm]{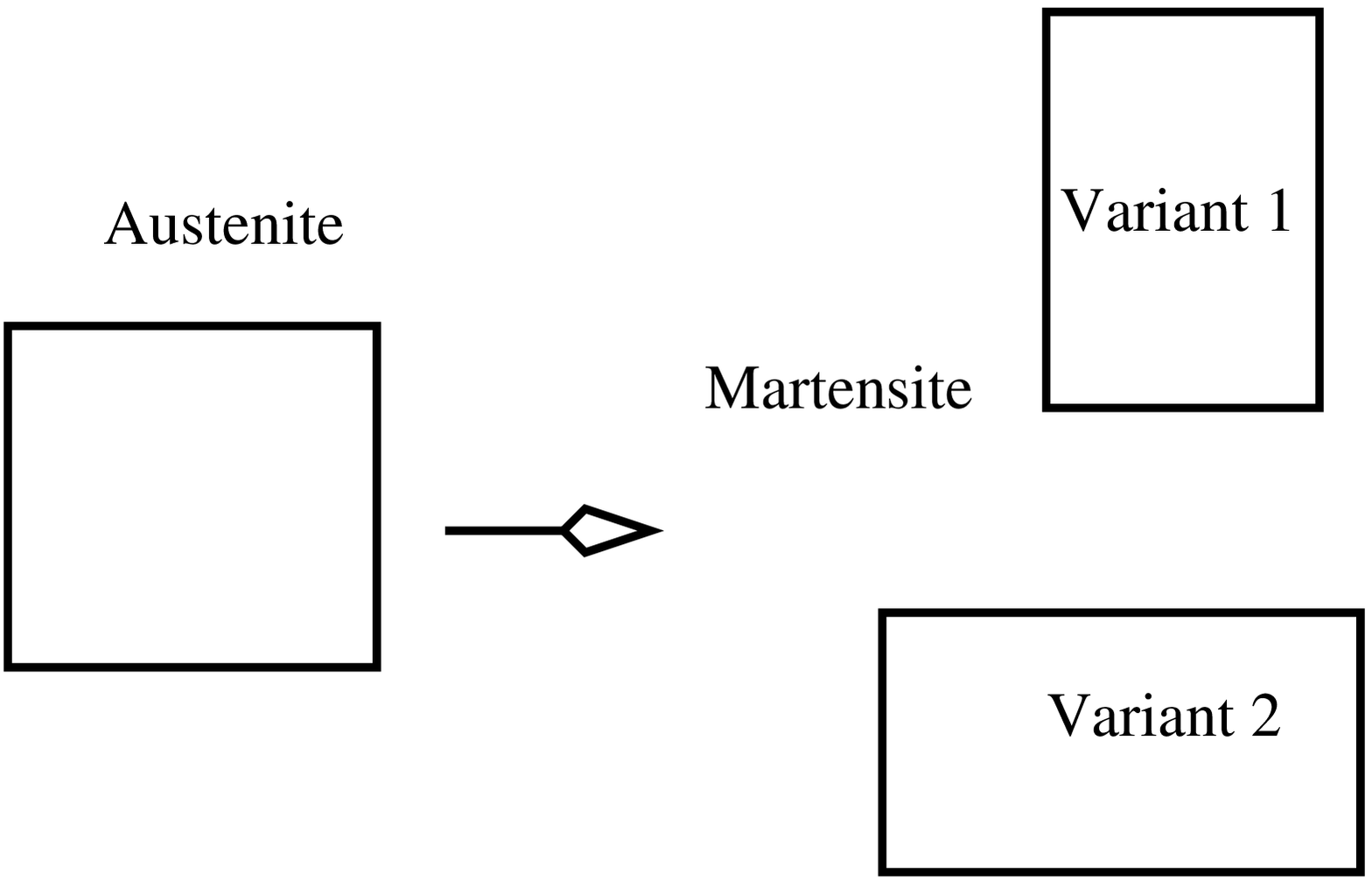}
        \vspace{4cm}
  \includegraphics[height=8cm, width=12cm]{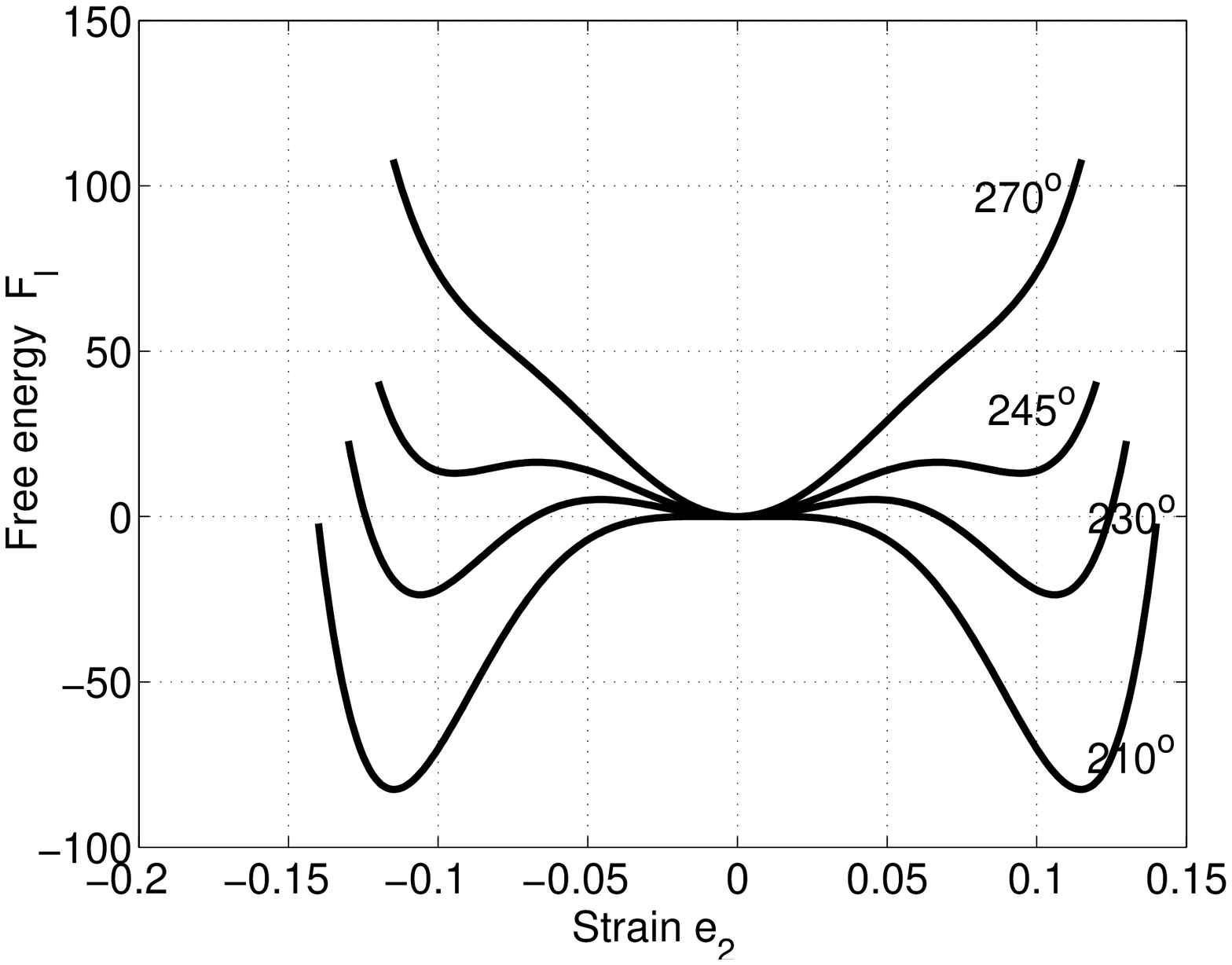}
  \caption{ (a) Sketch of the square to rectangular transformation, (b)
   The  temperature dependency of the free energy function for the
transformation. }
     \label{SRPT}
    \end{center}       \end{figure}

For the modeling of phase combinations in SMAs structures, the first task is to
characterize different phases, which is different from one material to another.
Here our mathematical model will be based on the square to rectangular
transformations. In  Fig.\ref{SRPT} (a) we give a schematic representation of
this case where the square lattice
is the austenite, while two rectangles are the martensite variants. The
square to rectangular transformation  could be regarded as a 2D analog of the
cubic to tetragonal or  tetragonal to orthorhombic transformations observed in
general 3D cases in $Nb_3Sn$, $InTl$, $FePd$ alloys and
some copper based SMAs (\cite{Jacobs2000,Lookman2003} and reference
therein).  The analysis of this  2D transformation  is a
first step in understanding more complex cubic to tetragonal and   tetragonal to
orthorhombic transformations.  Most numerical studies
of the dynamics of the phase transitions up to date  have been concentrated
on the analysis of the formation and
growth of the microstructure  (\cite{Lookman2003b,Jacobs2000,Lookman2003}
and references therein). Such studies have been focused on the
mesoscale under either periodic or affine boundary conditions.

In what follows, we base our consideration on the Landau theory of phase
transformation.  According to this theory, the basis of any nonlinear
continuum thermodynamical model for phase transformation is a non-convex {\it
free energy function} \cite{Melnik2000,Jacobs2000,Lookman2003}. The local minima of the
free energy function with respect to the strain tensor (or deformation
gradients) correspond to the stable and mesostable state at a given
temperature, while the microstructure in the domain of interest can be
described by the
minimizer of the bulk energy in the  domain. One of the simplest realization
of this idea in the context of SMAs is based on the Helmholtz free energy $\Psi$  (\cite{Falk1980,Melnik2000,Lookman2003,Matus2004}and references therein):
 \begin{equation} \label{eq2-1}
\Psi(\theta,\eps) = \psi_0(\theta) + \psi_1(\theta) \psi_2(\eps) +
\psi_3(\eps),
\end{equation}
\noindent where $\psi_0(\theta)$ models thermal field contributions,
$\psi_1(\theta)\psi_2(\eps)$ models shape memory contributions and
$\psi_3(\eps)$ models mechanical field contributions, $\eps=\partial{u}/\partial x $ ($u$ is the
displacement) is the strain which is chosen as the
only order parameter in the 1D case.  The  thermal field contributions $\psi_0$
can often be modelled as follows \cite{Falk1980,Wang2004} :
\begin{equation}
\label{eq2-2}
\psi_0 = -c_v \theta \ln \theta,
\end{equation}
\noindent  where $c_{v}$ is the specific heat constant.

The numerical analysis of a system of conservation laws based on the above
representation of the free energy function (and a more general one) has been
recently reported in detail in \cite{Matus2004}. A conservative numerical
scheme was constructed for the solution of the problem. It was noted
that a standard energy inequality technique, applied to the convergence
analysis of the scheme, can lead to quite restrictive assumptions. In
\cite{Matus2004} it was shown how such assumptions can be removed.

This work focuses on the practical development of an algorithm
suitable for the simulation of phase combinations. In this context we note
that a  free elastic energy functional (denoted further by $F$), similar to
the one discussed above, was established earlier
to characterize the austenite at high temperature and the martensite variants at 
low temperature in SMA patches, specifically for the square to rectangular 
transformation where the Landau free energy function $F_{l}$
was modified  \cite{Jacobs2000,Lookman2003,Lookman2003b,Wang2004}.
Recall that for the square to rectangular
transformation we have to deal only with two martensite variants and only one order 
parameter  \cite{Falk1980,Jacobs2000,Lookman2003} in order to characterize the
martensite variants and austenite in a 2D domain. Following previous works on
the subject (\cite{Falk1980,Jacobs2000,Lookman2003,Wang2004} and references
there in), we have:
\begin{equation}\begin{array}{l}
\displaystyle\label{eq2-3}
  F =   F_s  +  F_g ,
       \\[10pt] \displaystyle
 F_s  = \frac{a_1}{2}e_1^2+\frac{a_3}{2}e_3^2 +  F_l
      \\[10pt] \displaystyle
  F_l  =
\frac{A_2}{2}e_{2}^2 - \frac{a_4}{4}e_{2}^4+\frac{a_6}{6}e_{2}^6,
       \\[10pt] \displaystyle
 F_g = \frac{d_2}{2}\sum_{i=1}^{3} (\nabla
e_{i})^2 +   \frac{d_3}{4}\sum_{i=1}^{3} (\nabla^2 e_{i})^2 .
 \end{array}
\end{equation}
\noindent  where $\nabla$ is the gradient operator,
$A_2$, $a_{i}\quad i=1,\ldots, 6$,   $d_2$, and $d_3$ are material-specific
coefficients, and $e_{1}$, $e_{2}$, $e_{3}$ are dilatational, deviatoric, and shear
components of the strains, respectively,  defined as follows:
\begin{equation}
\label{eq2-4}
\begin{array}{l} \displaystyle
e_{1}=\left(\eta_{11}+\eta_{22}\right)/\sqrt{2},
     \\[10pt] \displaystyle
e_{2}=\left(\eta_{11}-\eta_{22}\right)/\sqrt{2},
        \\[10pt] \displaystyle
e_{3}=\left(\eta_{12}+\eta_{21}\right)/2.
\end{array}
\end{equation}
The Cauchy-Lagrangian strain tensor $\vec{\eta}$ is given by its
components
 \begin{equation}
\label{eq2-5}
\eta_{ij}\left(\textbf{x},t\right)=\left(\frac{\partial
u_{i}\left(\textrm{\textbf{x}},t\right)}{\partial x_{j}}+\frac{\partial
u_{j}\left(\textbf{x},t\right)}{\partial x_{i}}\right)/2,\kern0pt
\label{eq3}
\end{equation}
\noindent where $u_i$ is the displacement in the $i^{th}$ direction in the
Cartesian system of coordinates, \textbf{\textit{x}}$=(x_1, x_2, x_3)$ are the coordinates of a material point in
the domain of interest. It is known that in this formulation the deviatoric strain
$e_2$ can be chosen as the order parameter. More precisely, $e_2$ and $e_3$ are
two-component order parameter strains that have been discussed before in \cite{Lookman2003b,Ahluwalia2004,Ahluwalia2006}.

In the above  formulation the Ginzburg term
$ F_g$ is the term proportional to  square of strain gradients.
This term is often included to account for  the
presence of domain walls. It is essential in simulating phase growth and
several other phenomena (\cite{Jacobs2000,Lookman2003}and references
there in), but it can be ignored if we are interested only in the macroscopic phase combinations
of the SMAs patch under mechanical loadings (indeed,   the simulation scale in
this case is too coarse to
capture mesoscale structures).  Furthermore, the energy contribution of this
term is typically small compared to other terms.

In order to be able to model the entire range of different phase combinations
  that correspond to different temperatures,  
  the
material parameter $A_2$ is assumed  to be temperature dependent
$A_2=a_2(\theta-\theta_{0})$, where \textit{$\theta$}$_{0}$ is a critical
  temperature, responsible for the appearance of an addtional minimum and
  corresponding to the austenitic phase when temperature increases.  In Fig.\ref{SRPT} (b) we present the plots of
  the Landau free energy function, defined in this way, for the entire range of
  temperatures  of interest (the material
  $\textrm{Au}_{23}\textrm{Cu}_{30}\textrm{Zn}_{47}$). We observe
 that the function has two local minima at
low temperatures ($210^o$), which correspond to  two (rectangular in the
  interpretation of Fig.\ref{SRPT} (a))
martensite variants, while only one minimum at the center corresponds to the (square)
austenite phase when the temperature is high ($270^o$). When the temperature is in
between two critical values (e.g., at around $245^o$ in the figure), we
  observe that there are three local minima demonstrating co-existence
of metastable and stable phases.

Now, if we take the thermal contribution  $\psi_0$ the same as in the 1D case, the
final form of the Helmholtz free energy function for the square to rectangular
transformation will take the following form:
\begin{equation}
\label{eq2-6}
\Psi(\theta,\eps)=-c_{v} \theta \ln \theta +
\frac{a_{1}}{2} e_{1}^{2}+ \frac{a_{3}}{2}e_{3}^{2}+F_{L},\quad
F_{L}=\frac{a_{2}}{2}\left(\theta-\theta_{0}\right)e_{2}^{2}-\frac{a_{4}}{4}
e_{2}^{4}+\frac{a_{6}}{6}e_{2}^{6}.
\end{equation}

\noindent By substituting the free energy function into the variational problem given
by Eq.(\ref{eq1-1}), the phase combination problem can be written as the following
variational problem. Given  temperature $\theta$, find the
displacements $u_x$ and $u_y$ (in the $x$ and $y$ direction)  that minimize the
bulk energy:
\begin{equation}
 W(u_x,u_y )=\int_{\Omega} \left ( -c_{v} \theta \ln \theta +
\frac{a_{1}}{2} e_{1}^{2}+ \frac{a_{3}}{2}e_{3}^{2} +
\frac{a_{2}}{2}\left(\theta-\theta_{0}\right)e_{2}^{2}-\frac{a_{4}}{4}
e_{2}^{4}+\frac{a_{6}}{6}e_{2}^{6} \right ) dV.
\label{eq2-7}
\end{equation}

Under a given temperature, the contribution of the thermal field will not 
change the profile of the local free energy
function, but rather only shift it upwards
or downwards.  Hence, if the applied external force in 2D is given by
 its components ${f_x, f_y}$,  the final problem to solve can be formulated as
follows:
\begin{equation}
 \begin{array}{l}
 W(u_x,u_y)=   \\    \displaystyle     
      \int_{\Omega} \left (  
\frac{a_{1}}{2} e_{1}^{2}+ \frac{a_{3}}{2}e_{3}^{2} +
\frac{a_{2}}{2}\left(\theta-\theta_{0}\right)e_{2}^{2}-\frac{a_{4}}{4} 
e_{2}^{4}+\frac{a_{6}}{6}e_{2}^{6}- f_x u_x - f_y u_y \right  ) dV \rightarrow
      \min. 
\end{array}
\label{eq2-8}
\end{equation}
 
The above model is reduced to the well-known Falk model in the 1D case
\cite{Falk1980} that can be written with respect the only order parameter $\epsilon= \partial u / \partial x$:
   \begin{equation}
   W(u)=\int_{\Omega} \left (
   \frac{a_{2}}{2}\left(\theta-\theta_{0}\right)\epsilon^{2}-\frac{a_{4}}{4}
\epsilon^{4}+\frac{a_{6}}{6}\epsilon^{6} - f u \right ) dV,
   \label{eq2-9}
   \end{equation}

\begin{figure}      \begin{center}
      \includegraphics[width=7cm, height=6cm]{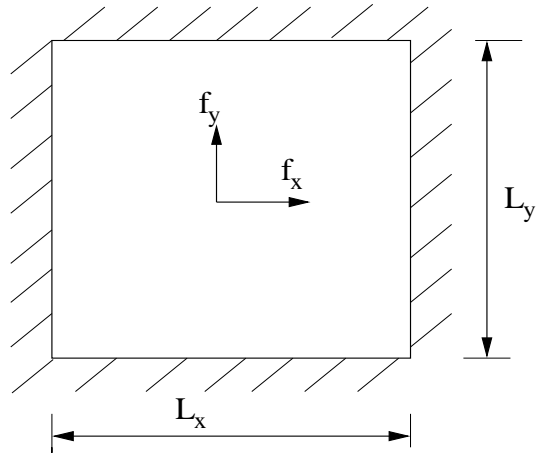}
       \caption{Sketch of a two dimensional SMA patch under mechanical loading}
    \label{SMA2D}
      \end{center}            \end{figure}

As discussed in Section 1, we supplement the model (\ref{eq2-8}) by
appropriate boundary conditions. For all examples discussed in Section 4,
these are  clamped boundary conditions:
      \begin{equation}    u_x=u_y = 0, ~ \textrm{at} ~  x = x_l,
x_r, ~\textrm{or}~      y= y_t,y_b,
     \end{equation}
where  $x_l$ and $x_r$ are the left and right boundaries along the $x$ direction,
$y_t$ and $y_b$ are the top and bottom boundaries along the $y$ direction, as sketched in
Fig.\ref{SMA2D}. External forces vary and are specified in Section 4.


\section{Numerical Implementation Based on Hybrid Optimization}

The above variational problem is non-convex and its solution  can
only be obtained by numerical methods. The procedure developed in this section consists
of two main steps: firstly, the
variational problem is  converted into a nonlinear
minimization problem by spatial discretization, and then the solution of the
resulting problem is sought by a hybrid optimization
strategy.

\subsection{Spatial Discretization Procedure}

For the spatial discretization, we employ the Chebyshev pseudospectral
approximation (e.g., \cite{Alfio2000}) on a set of 2D Chebyshev points $(x_i, y_j )$ in the 2D domain
of interest  $\Omega = [-1,1] \times [-1,1]$:    
\begin{equation}
  \label{eq3-1}
  x_i  =  \cos(\frac{\pi i}{N}) ,
  y_j  =  \cos(\frac{\pi j}{N}) ,
   \quad i,j=0,1,\dots,N,
\end{equation}
\noindent where $N+1$ is the total number of nodes in one direction.
Other structures of interest
can be mapped onto the domain $\Omega$ by a
linear transformation. 
 Using the constructed grid, the
displacements in the patch can be approximated as follows:
\begin{equation}
  \label{eq3-2}
    f(x,y) = \sum_{i=0}^{N} \sum_{j=0}^{N} f_{i,j}
   \xi_i(x)\xi_j(y),
\label{3eq2}
\end{equation}
\noindent where $f(x,y)$ represents either  function $u_x$ or $u_y$, $f_{i,j}$ is
the function value at $(x_i,y_j)$. Functions $\xi_i(x)$ and $\xi_j(y)$  are the $i^{th}$
and   $j^{th}$ Lagrange interpolating polynomials along the $x$ and $y$ directions
respectively.

Having obtained (\ref{3eq2}), the derivatives of
function $f$, $\partial f(x,y)/ \partial x$ and $\partial f(x,y)/ \partial y$, can be
obtained by calculating $\partial \xi_i(x)/ \partial x$  and $\partial
\xi_j(y)/ \partial y$. Following the standard technique found, e.g., in
\cite{Trefethen2000,Alfio2000}, all the differentiation operators in
Eq.(\ref{eq2-8}) (or Eq.\ref{eq2-9}) can be written in the matrix forms:
\begin{equation}
 \vec F_x =  \vec D_x \vec F
, \quad
 \vec F_y =  \vec D_y \vec  F,
\label{eq3-3}
 \end{equation}
\noindent where $\vec
F_x$ and $\vec F$ are vectors collecting all values of the derivative $\partial f
/\partial x$ and the function $f$ at $(x_i, y_j)$, respectively, and similarly for $F_y$.
 The differentiation matrices $ \vec D_x$ and $\vec D_y$
 can be calculated using the approximation given by Eq.(\ref{eq3-2}). For instance,  the
differentiation matrix $ \vec D_x$ for the Falk model takes the following form:
\begin{equation}   \label{eq3-4} \vec D_{ij} = \left \{
 \begin{array}{ll}   \displaystyle
  \frac{2N^2+1}{6} &  i=j=0,
       \\[10pt] \displaystyle
  -\frac{2N^2+1}{6} &  i=j=N,
       \\[10pt] \displaystyle
  -\frac{x_j}{2(1-x_j^2)} &  i=j=1,2,\dots, N-1,
       \\[10pt] \displaystyle
   \frac{c_i}{c_j} \frac{(-1)^{i+j}}{(x_i-x_j)} &
      i \neq j,\quad i,j=1,2,\dots, N-1,
  \end{array}
  \right.
\end{equation}
\noindent  where $c_i=2$ for $i=0, N$ and $c_i=1$ otherwise.
Such matrices have dimensionality $(N+1) \times (N+1)$.

The  bulk energy is given by an integral operator with  the local free energy
function as its integrand. We use the same set of points as chosen for the
derivative approximation for constructing a numerical
integration formula for integral. In particular, we use  Chebyshev collocation
nodes and the resulting
quadrature formula is constructed by using the Chebyshev-Lobatto rule
\cite{Alfio2000,Trefethen2000}. For example, the formula for integration in the x-direction, generically represented by  
  \begin{equation}
      \int_0^1 f(x) dx \approx \sum_{i=0}^N w_i f(x_i),
\label{3eq19}
    \end{equation}
 is exact for any polynomials with an order less
than $2N-1$.
In (\ref{3eq19}) weight coefficients $w_i$ are defined in the standard manner \cite{Alfio2000,Gautschi1999}.
  
By substituting the approximations of all the differentiation operators
and integral operators into the variational problem,  we convert the original problem into the following minimization problem:
 \begin{equation}
      \textrm{find}~ u_{i,j}^1, u_{i,j}^2 ~\textrm{to minimize:}~
       W(u_{i,j}^1, u_{i,j}^2),
  \end{equation}
 \noindent  where  $u_{i,j}^1$  and $u_{i,j}^2$   stand for the values of $u_x$
and $u_y$ at node $(x_i, y_j)$, respectively.   Note that $W(u_{i,j}^1, u_{i,j}^2)$ is 
just the discretized bulk energy and is a nonlinear algebraic function of 
$u_{i,j}^1$ and $u_{i,j}^2$.  The resulting problem has  $2\times(N+1) \times (N+1)$
variables in total.  Given the prescribed boundary conditions,  we have $2\times (N-1) \times
(N-1)$  variables in total.
 
The  problem to be solved is non-convex minimization problem with strong
nonlinearity. It is not easily amenable to conventional gradient-based minimization
methodologies due to multiple local minima. On the other hand, the genetic
algorithms (GA) can be helpful in locating an approximation to the global minimum,
giving an initial approximation to gradient-based procedures. In what follows,
we combine these two ideas by employing a hybrid optimization strategy that
takes advantage of the
global exploring capability of  the GA  and the local refinement accuracy of the
quasi-Newton method for non-convex problems.

 \subsection{Genetic Algorithm Locates Initial Approximations}

The GA is a well established methodology for global optimization (\cite{Goldberg1989,Mitchell1996,Forrest1993} and references therein). We highlight
here only its main features in the context of our problem.
The GA
maintains a population of individuals  (chromosomes), say $P(n)$, for generation $n$ and
each chromosome consists of a set of genes, where each gene stands for a parameter to be
estimated. One chromosome represents one potential solution to the minimization problem.
Each chromosome is evaluated to give  some measure of its fitness according to the
bulk energy defined in the previous section. Some chromosomes undergo stochastic transformations by means of
genetic operations to form new chromosomes. Recall that there are two transformations in the GA:
{\it crossover}, which creates new chromosome by combining parts from two chromosomes, and {\it mutation}, which creates a new chromosome by making changes in a single
chromosome. New chromosomes, called offsprings $S(n)$, are then evaluated. A new
population is formed by selecting fitter chromosomes from the parent population
and the offspring population. After some generations, the algorithm converges to
the fittest chromosome, which  represents an estimated optimal  solution to the
problem \cite{Goldberg1989,Mitchell1996,Forrest1993}.

{\it Generation of initial chromosomes}: In most of the GAs, the chromosomes are
created by randomly choosing  the genes in a given range. For the current
problem, this is not an effective way.  Indeed,  if the strain in
solid structures under consideration is assumed to be not very large, while
displacements  may vary 
slowly and be represented by smooth functions, there is no good reason to
include high frequency oscillations in 
displacements.  At the same time, if the displacement values are
randomly chosen on our discrete grid,  high frequency oscillations may well be
pronounced.   Hence, we smooth the randomly chosen chromosomes by the  following filter operation:
 \begin{equation}   \vec U_s=   \vec I_d  \vec I_{s}   \vec  U_r,
\end{equation}
\noindent where $ \vec U_r$ is the vector collecting  randomly chosen  displacement
values at all discretization nodes, while $  \vec U_s$ is the resultant smoother
profile of the displacements after the filter operation. The matrix $\vec
I_s$ is an interpolation matrix  which maps  the given data at the discretization
nodes onto a sparser grid, using the least square approximation, while the
matrix $\vec I_d$ is another interpolation matrix which maps the data on the sparse
grid back to the given discretization nodes.  These two matrices can be
constructed by using the Chebyshev collocation method, as discussed above.  The
product of these two interpolation matrices could be regarded as a filter to
remove high frequency components.

{\it Crossover}: Crossover operator in the GA is for producing new children
chromosomes from chosen parent chromosomes. For a given pair of parent
chromosomes $\vec x_1$ and $\vec x_2$, the offspring is obtained by the so
called Intermediate recombination \cite{Mitchell1996,Gautschi1999}:
\begin{equation}
  \vec o_1 = \vec  x_1 \vec  \alpha + \vec x_2 (1 - \vec \alpha),
\end{equation}
\noindent  where  $\vec \alpha$ is a vector with the same size as $\vec x$, and
all its entries are randomly chosen independently.  This operation is  capable
of producing variables slightly larger than the hypercube in the solution space
defined by the parents, but confined by the parameter $\vec \alpha$. A typical
 range for the parameters in  $\vec \alpha$ is
$[-0.25, 1.25]$, which is used in the current paper.

{\it Mutation}: The mutation operation is randomly applied with a low
probability, typically in the range  $0.001$ to $0.01$, and modifies genes in
the chosen chromosome.  Its role in the GA is to make sure that the probability
of searching any potential solution (any points in the solution space) is
nonzero, and is often regarded as a measure  to recover good genetic material
that might  be lost through the operation of  selection and crossover
\cite{Mitchell1996,Goldberg1989,Forrest1993,Gautschi1999}.

In practice, the mutation operation is carried out by replacing a randomly
chosen gene by a randomly chosen new value in the given range. The mutated
chromosome itself is randomly chosen from the current population with a given
probability. In the current problem, the mutation operation is applied to one
chromosome in one generation, which means the probability for each chromosome is one
divided by the number of chromosomes.

{\it Generation-alteration method}:  Generation alteration method determines how 
to evolve the current generation to the next, which means how to select pairs of
parents for producing  children by crossover and mutation operator, and how to 
select parents in the currents population that survive in the next generation.

Taking into account their fitness, it is obvious that each chromosome should have
a probability determined by the associated function value. Here a simple linear map method
is used as follows:  rank all the chromosomes in the current  generation in order
of decreasing function values, then the probability of being chosen for a
specific chromosome $i$ is calculated as:
\begin{equation}
\rho_i= \frac{i}{2(N+1)N},
\end{equation}
\noindent where $i$ is the position of the chromosome in the rank, while $N$ is the number of
chromosomes in the rank.  Then the parents for survival and crossover are chosen
 using the above calculated probability.  While 
all the chromosomes have nonzero probability to be chosen, fitter chromosomes
have a better chance.

The results of application of this procedure to the simulation of SMA
phase combinations are reported in the next section.

\subsection{Quasi-Newton Method Refines the Solution}

The output of the GA is used as  the initial guess for local search methods.
In what follows we apply the quasi-Newton method  to the minimization of the
bulk energy given in Eq.\ref{eq2-8}.  Let's denote any potential
minimizer  by $\vec x$.  The bulk energy   $W(\vec x )$ is  minimized by
$\vec  x^*$  if it satisfies   $\nabla W(\vec x^*)= 0$. To achieve this, we can
organize the following iterative process. Let at  the general $k^{th}$ step of the iteration, the potential solution is $\vec
x_k$.  Then, the task is to estimate  the next $\vec x_{k+1}=\vec
x_k + \vec d_k$ such that   $  \nabla W(\vec x_{k+1} ) = 0$ which means that
     \begin{equation}
      \nabla W(\vec x_k + \vec d_k ) = \vec q(\vec x_k + \vec d_k) =
       \vec q(\vec x_k) +  \nabla \vec q (\vec x_k ) \vec d_k = 0,
   \end{equation}
\noindent  where  $\vec q (\vec x_k) =  \nabla W(\vec x_k) $ is the gradient of
the bulk energy at $\vec x_k$,    $\nabla \vec q (\vec x_k ) =\nabla^2 W(x_k) $
is the Hessian matrix,  while   $\vec d_k$ is the search direction. To make $  \nabla
W(\vec x_{k+1} ) = 0$, the search direction should satisfy:
     \begin{equation}
      \vec d_k  =  -  (\nabla \vec q (\vec x_k ) )^{-1} \vec q(\vec x_k).
       \label{quasiNewton}
     \end{equation}
This standard Newton-type procedure is computationally expensive and has other
well-known 
drawbacks \cite{Alfio2000,Yuan1997} that preclude us from using it in the
context of our problem. Instead, we apply the quasi-Newton procedure by constructing a matrix $\vec B_k$ (at the $k^{th}$ iteration), an
approximation to the Hessian matrix that  satisfies the following condition:
   \begin{equation}
         \vec q(\vec x_k) +     \vec B_k  \vec d_k = 0.
  \end{equation}

Provided with the initial guess from the GA, the local search method using 
the quasi-Newton method is organized in a standard manner: at the 
general $k^{th}$ iteration  ($\vec x_k$ is the current estimated solution):  
     \begin{itemize}
  \item[(a)]  Find a descent direction $\vec d^k$ using $ \vec B_k$  by the following
  formula:
            \begin{equation}
                \vec d^k = - \vec  B_{k}^{-1} \nabla  W(\vec x^k),
              \end{equation}
 \item[(b)]  Compute the acceleration parameter $\alpha_k$ by line search,
  find $\alpha_k$ such that $W( \vec x^k+\alpha_k \vec d^k)$ is minimized.
 \item[(c)]  Update the potential solution:
     \begin{equation}
            \vec   x^{k+1} = \vec x^k + \alpha_k d^k .
      \end{equation} 
\end{itemize}

Finally,  the update of the approximation of the Hessian matrix $\vec B_{k+1}$ 
is organized by using the   BFGS update
(Broyden-Fletcher-Goldfarb-Shanno update, e.g., \cite{Alfio2000,Yuan1997,Martinez2000}):

 \begin{equation}
   \vec  B_{k+1} =\vec  B_k -  \frac{(\vec {B_kd_k}) (\vec {B_kd_k})^T}
          {\vec d_k^T \vec B_k \vec d_k}
        + \frac{\vec y_k \vec y_k^T}{\vec y_k^T d_k},
     \end{equation}
where $\vec y_k=\nabla W(x_{k+1}) - \nabla W(x_{k}) $.
 To initiate the iteration process, $B_0$ for the first step is taken 
as the identity matrix that corresponds to the deepest descent methodology.
 For the current problem, we did not apply the line search for the acceleration parameter
$\alpha_k$. Instead, a  small value, regarded as a relaxation factor,  was assigned to $\alpha_k$.


\section{Numerical Examples: Phase Combinations in SMA with Hybrid
  Optimization Procedure}

By combining  the  GA and  the quasi-Newton method,
the bulk energy given in Eq.(\ref{eq2-8}) and Eq.(\ref{eq2-9}) can be
minimized with respect to displacements. To demonstrate the capability of this hybrid
optimization method, three numerical experiments are reported here. For all three  experiments, the GA first evolves a given number of generations,
and the fittest chromosome in the last generation is selected. This
fittest chromosome is then used as the initial guess for  the quasi-Newton method,
and is refined iteratively.  The termination criterion for the quasi-Newton
method is based on the assumption that the norm  of difference between 
two consecutive potential solutions is smaller than the predefined value
$\delta$ (chosen  in experiments as $\delta=1\times 10^{-6}$):
    \begin{equation}
   \Vert  s_{k+1} \Vert = \Vert \vec x^{k+1} -\vec x^{k}  \Vert    \leq \delta.
      \end{equation}

 All simulations reported here have  been carried out for
$\textrm{Au}_{23}\textrm{Cu}_{30}\textrm{Zn}_{47}$.  For this specific material,
its physical parameters are available  for the 1D case:
\cite{Falk1980,Niezgodka1991}:
 \begin{displaymath}
\begin{array}{c} \displaystyle
 a_{2}=480\, g/ms^{2}cmK,\qquad    a_{4}=6\times10^{6}g/ms^{2}cmK,
 \qquad a_{6}=4.5\times10^{8}g/ms^{2}cmK,
    \\[10pt]\displaystyle
\theta_{0}=208K,\quad  \rho=11.1g/cm{}^{3},\quad
  c_{v}=3.1274g/ms^{2}cmK, \quad k=1.9\times10^{-2}cmg/ms^{3}K.
\nonumber
\end{array}
\end{displaymath}

Since 2D experimental values are not available to us,  we take all the parameters in the Landau free
energy function the same as above and complete parameterization of the model by
assuming $a_1=2a_2$ and $a_3 = a_2$ as suggested in
\cite{Jacobs2000,Curnoe2001}. Earlier, we confirmed numerically  that the
essential features of the 2D problem can be captured with this
parameterization, at least in the case of  square to
rectangular transformations considered here \cite{Wang2004,Wang2004b}. Recent
 studies presented in \cite{Ahluwalia2006}  provided encouraging results also
  for more general cubic to tetragonal
transformations. However, experimental results on multi-dimensional SMA samples are
still lacking. The next step and a natural development of the study presented here
would be accounting systematically for the dynamics of the thermal field. Numerical
experiments pertinent to this generalization would be much more involved.

\begin{figure}    \begin{center}
    \includegraphics[width=10cm, height=3cm]{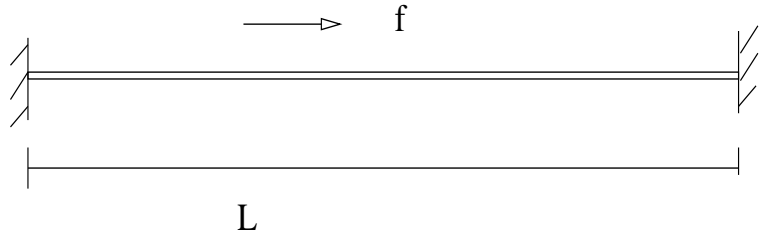}
     \caption{Sketch of an one dimensional SMA wire under mechanical loading}
        \label{SMA1D}
     \end{center}     \end{figure}

The first experiment is the simulation of phase combinations in a 1D
SMA wire with the bulk energy given by Eq.(\ref{eq2-9}). The physical
 interpretation of the problem is sketched in Fig.\ref{SMA1D}.  The length of the wire is
$1cm$ , and the applied mechanical force is  $f=500
g/\left(ms^{2}cm^{2}\right)$ which is evenly distributed along the whole
length. The initial condition for this case, $u=({\rm rand}-0.5)/5$, provides
a random distribution of displacement in the interval $[-0.1, 0.1]$.
  $15$ nodes are used for the spatial discretization. For the filter operation in the GA,
$7$ nodes are used for  the sparser grid  to remove the higher frequency components.
 The GA evolves $800$ generations with $60$ chromosomes in each generation,  each
chromosome consists of displacement values within the range from
$-0.1$  to $0.1$ on internal nodes.

\begin{figure}      \begin{center}
    \includegraphics[width=7cm, height=6cm]{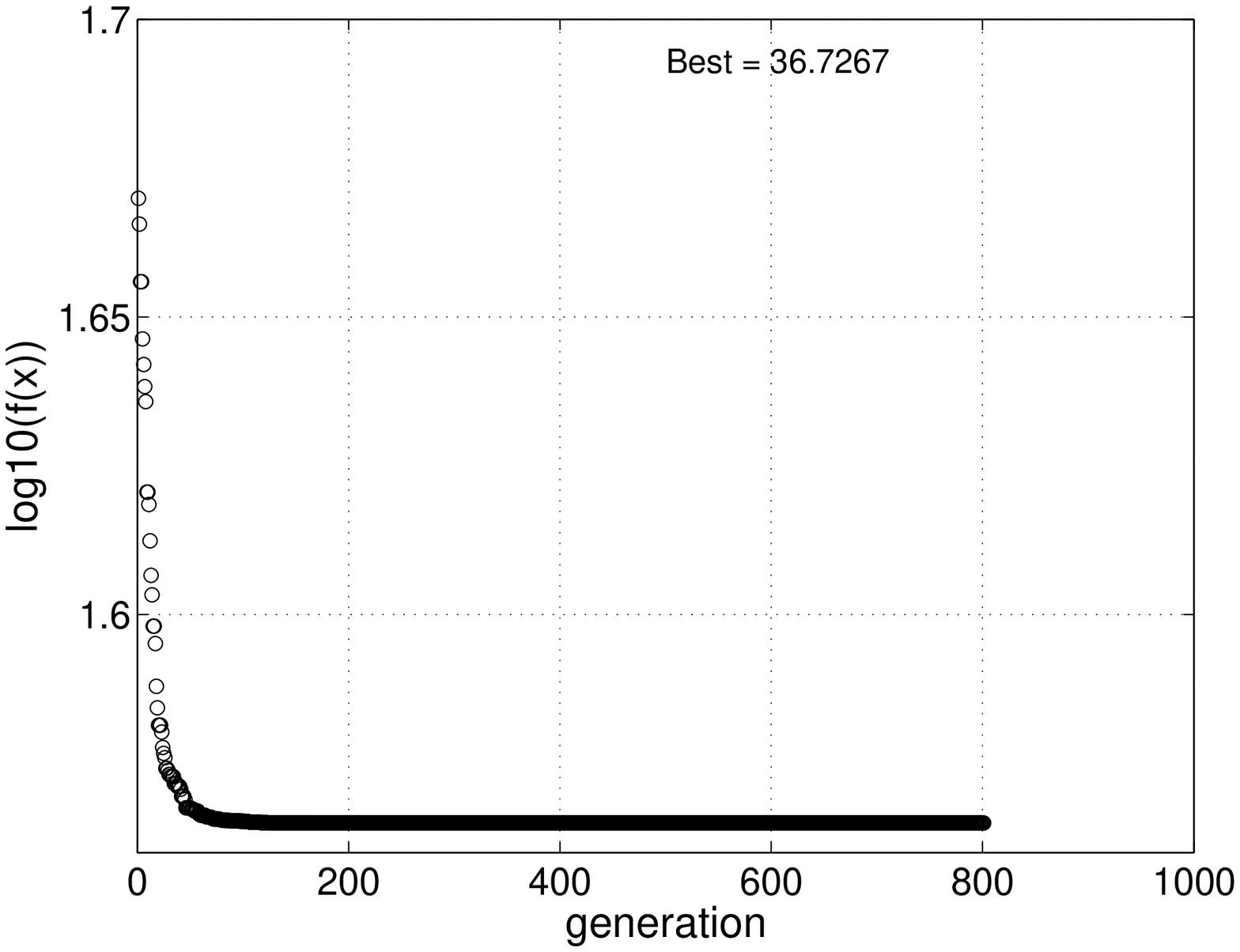}
    \includegraphics[width=7cm, height=6cm]{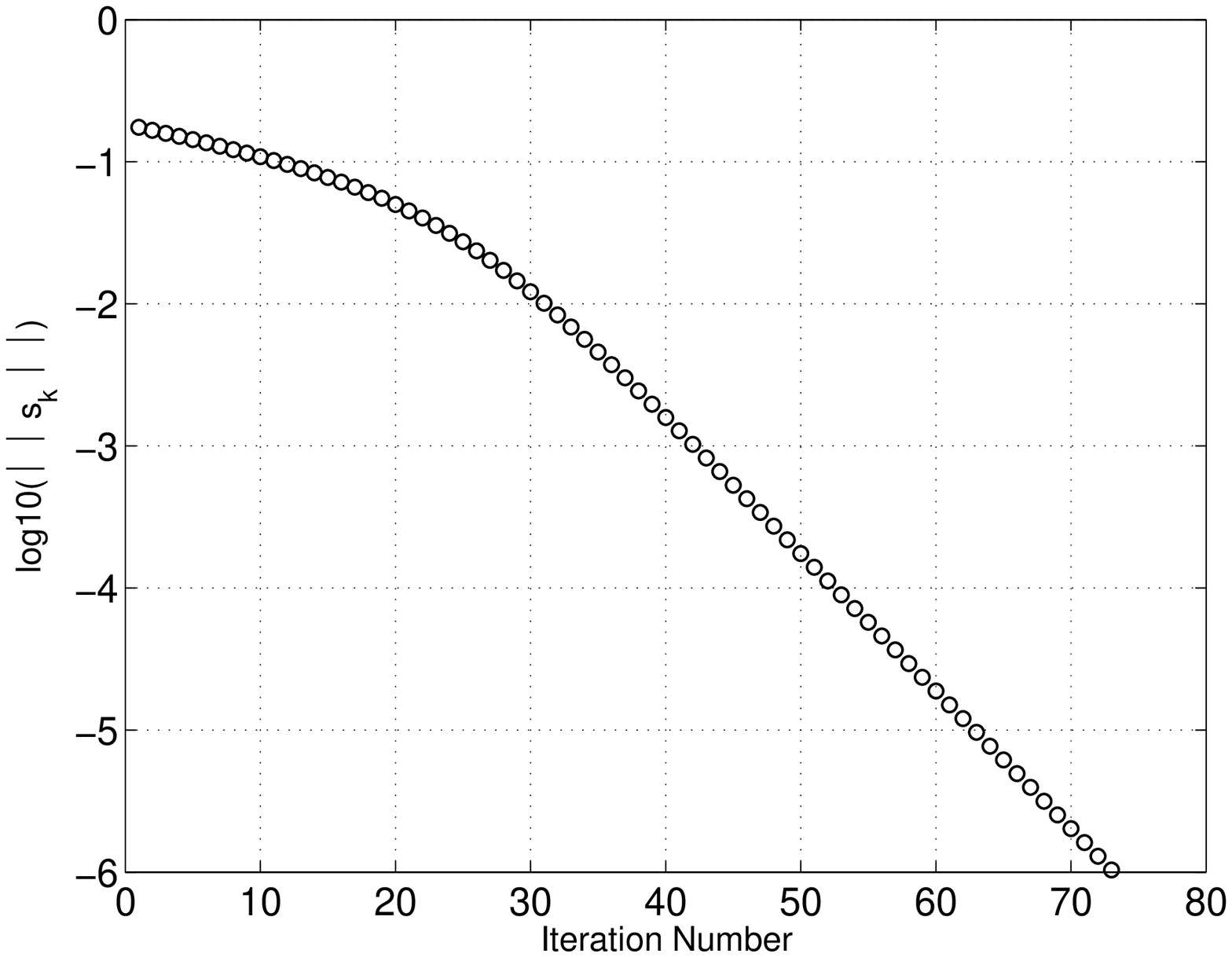}
     \caption{Convergence of the hybrid optimization method. Left: reduction of the bulk
energy in the GA. Right: reduction of the step size in the quasi-Newton method  }
      \label{Residual}
      \end{center}        \end{figure}

To analyze the performance of the hybrid optimization method for phase combination
simulation,  the bulk energy given by Eq.(\ref{eq2-9}) has been monitored in
the GA, and plotted in Fig.\ref{Residual} (left). The quasi-Newton iteration process has also been
monitored. The actual update step size in the
quasi-Newton method is plotted in Fig.\ref{Residual}  (right).
The trends in the two curves demonstrate how the  optimization process
evolves. As expected, the GA starts to converge to the global minimizer gradually, but has
difficulty to locate it precisely.  The quasi-Newton method starts with the output of the
GA as its initial guess, and refined the global minimizer effectively.

\begin{figure}        \begin{center}
  \includegraphics[width=7cm, height=7cm]{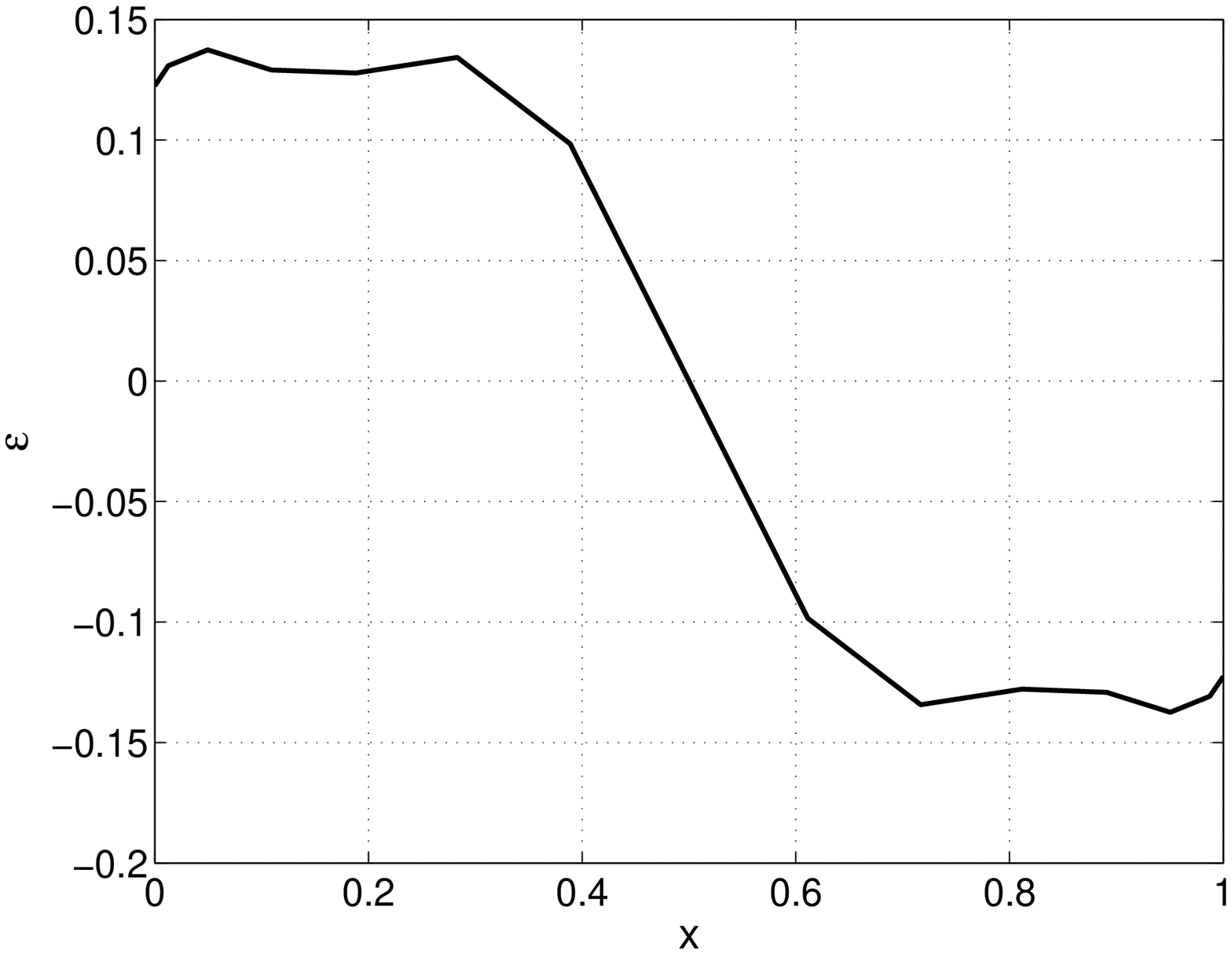}
  \includegraphics[width=7cm, height=7cm]{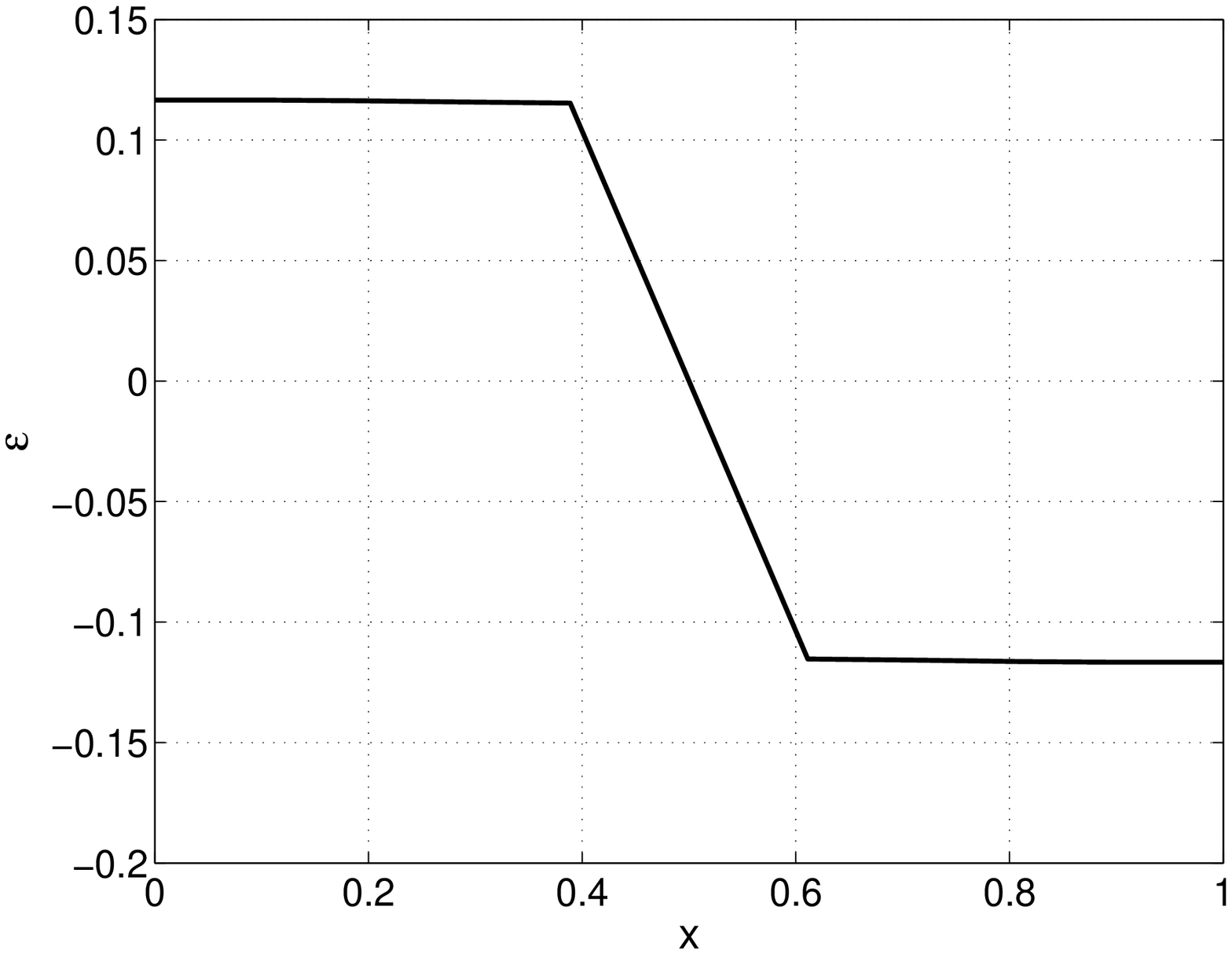}
  \includegraphics[width=7cm, height=7cm]{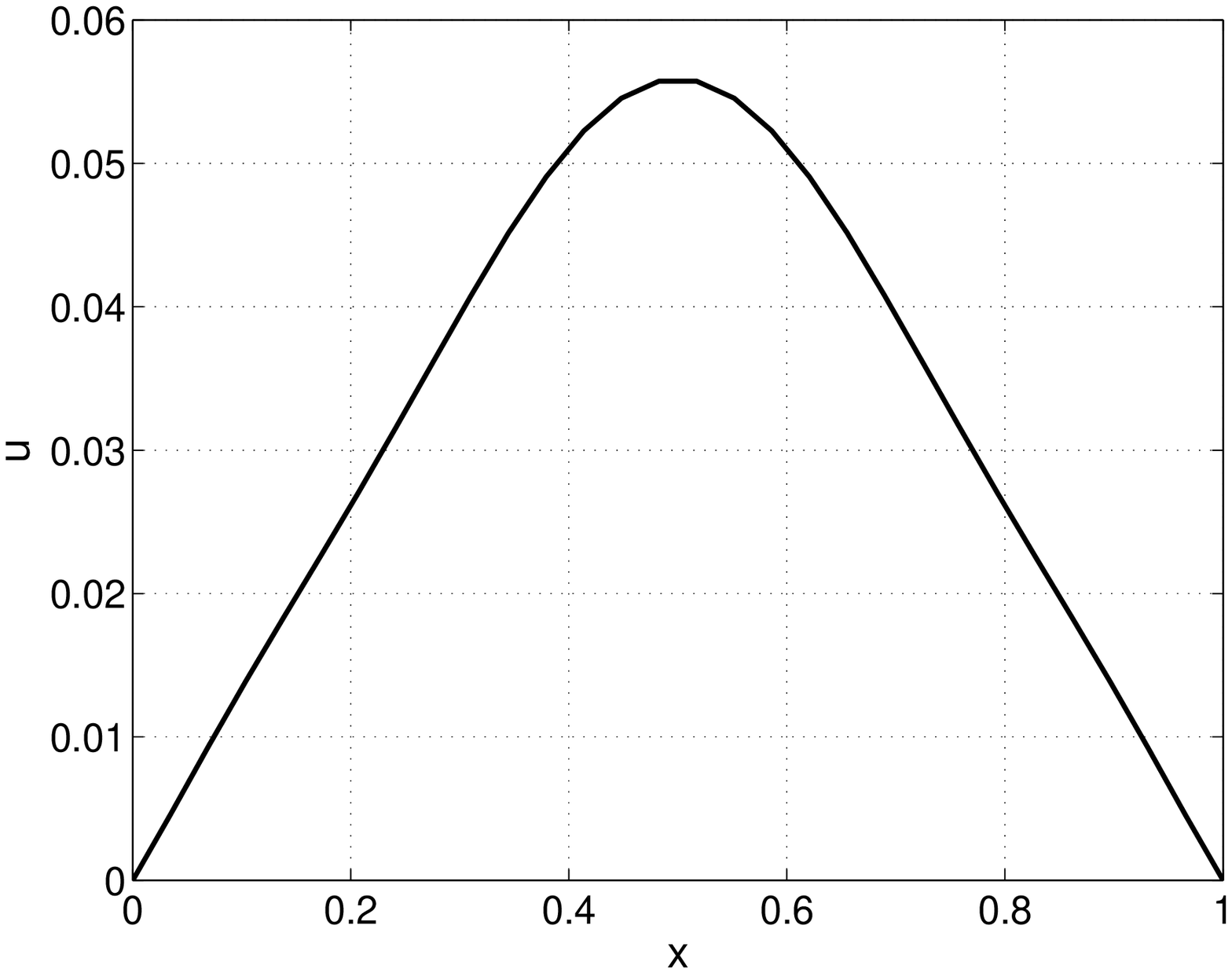}
  \includegraphics[width=7cm, height=7cm]{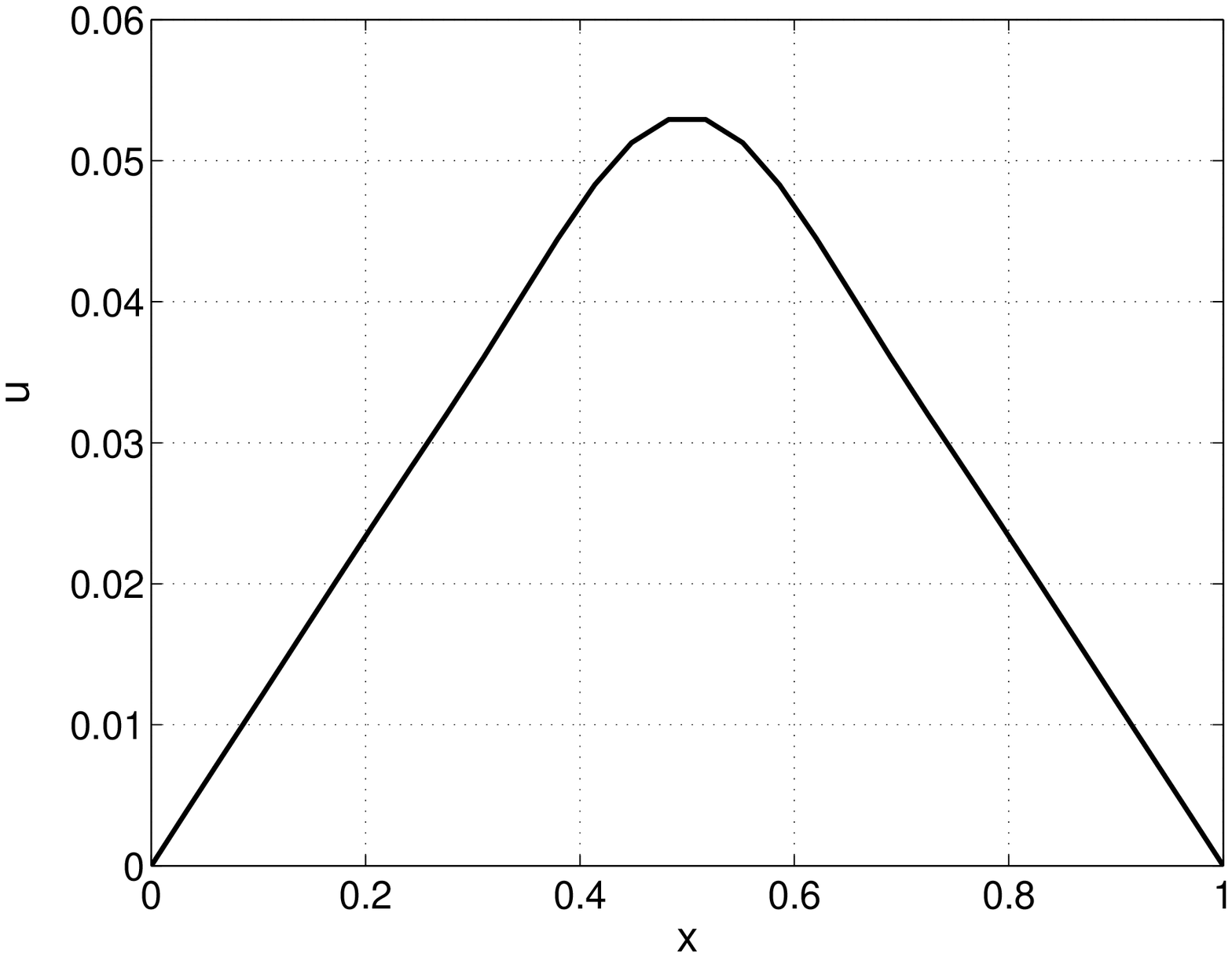}
  \caption{Mechanically induced phase combination in a SMA wire.}
 \label{Case1}
   \end{center}          \end{figure}

The final estimation for the phase combination in this case is given in  the right column
in Fig.\ref{Case1}: the order parameter $\epsilon$ (upper plot) and the displacement
(lower plot).  We observe that the entire domain is
divided into two parts, one with the strain values $\epsilon \simeq 0.11$ and
 the other with $\epsilon \simeq -0.11$. This phase combination agrees well
 with the previously reported  results  (e.g.,
 \cite{Falk1980,Luskin1996,Niezgodka1991}). For comparison purpose, we 
have also provided an approximation obtained with the GA at the initial stage
of the procedure (see the left column of Fig.\ref{Case1}). Although the
quality of the final solution has been refined with the quasi-Newton
procedure, we note that the two parts in the distribution of $\epsilon$ are
captured by the GA. Quantitative values are easily identified as being between
$0.1$ to $0.15$ for one of the domains and between  $-0.15$ to $-0.1$ for the other.

The second experiment aims at simulating the phase combination in a 2D SMA patch
 sketched in Fig.\ref{SMA2D}. The size of the patch  is $[-1, 1]\times [-1,1]
cm^2$ and the applied  mechanical forces  $f_x=f_y= 3000
g/\left(ms^{2}cm^{2}\right)$ are distributed evenly through the entire patch. 
In this case, we use 
 $u_x=({\rm rand}-0.5)/5$ and  $u_y=({\rm rand}-0.5)/5$ as the initial
 conditions. In this experiment,
we use different node numbers for the GA and the quasi-Newton method.  In the GA,
there are $9$ nodes in each direction used for the discretization along the
 $x$ and $y$ axes. This leads to the total number of optimization parameters
 being  $128$. The filter operation is
 carried out in the same way as explained for  the 1D experiment, but for both $x$ and $y$
directions,  and the number of nodes in each direction on the sparse grid is  $6$.  The GA is firstly run
$1500$ generations with $120$ chromosomes in each generation.  The output of the GA is
then interpolated onto a denser grid with $15$ nodes in each direction, and is
 used as
the initial guess for the quasi-Newton method. The interpolation methodology
 is based  on the pseudospectral method with the Chebyshev collocation points,
 as discussed in Section 3.

\begin{figure}   \begin{center}
  \includegraphics[width=7cm, height=7cm]{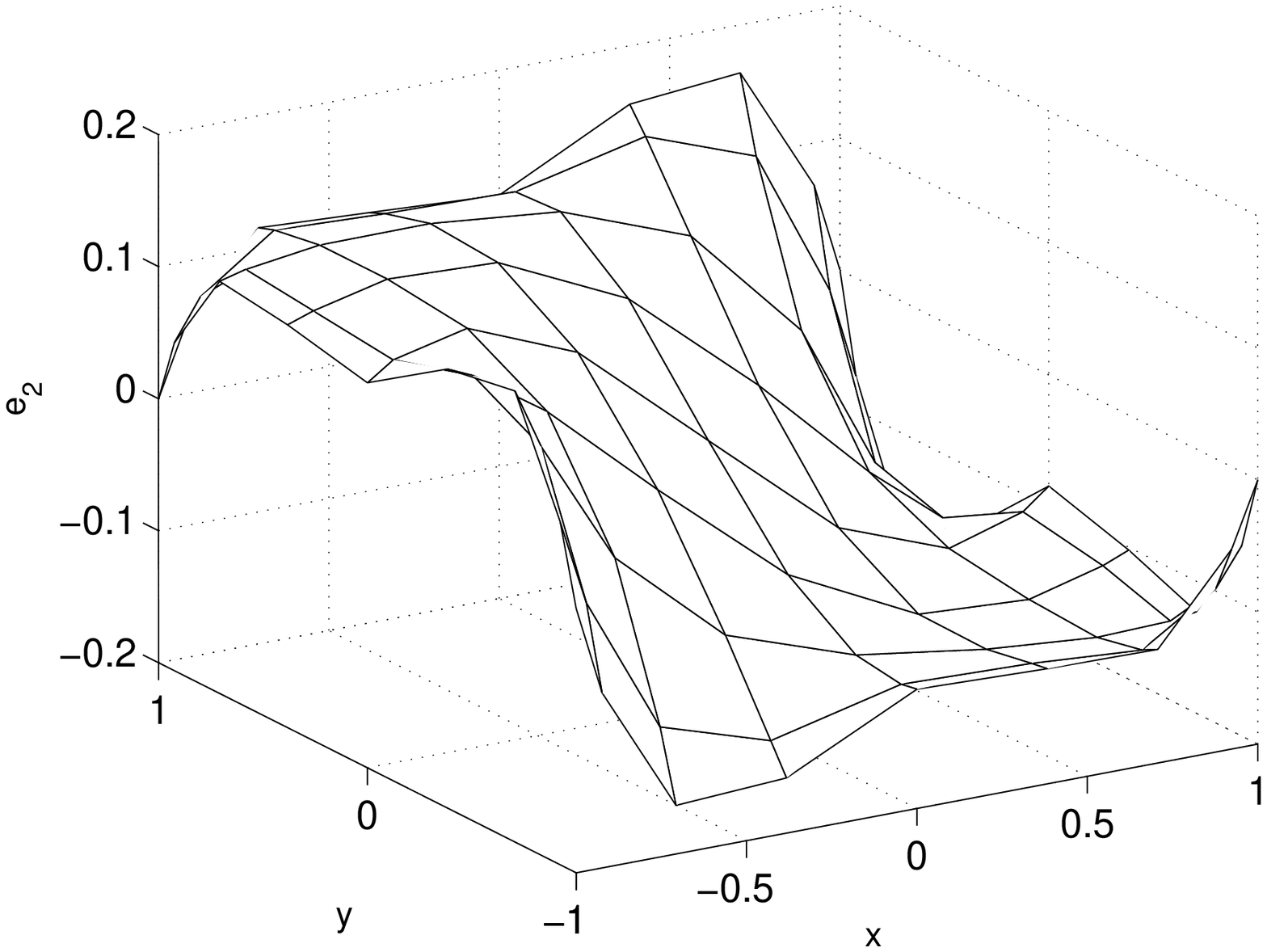}
  \includegraphics[width=7cm, height=7cm]{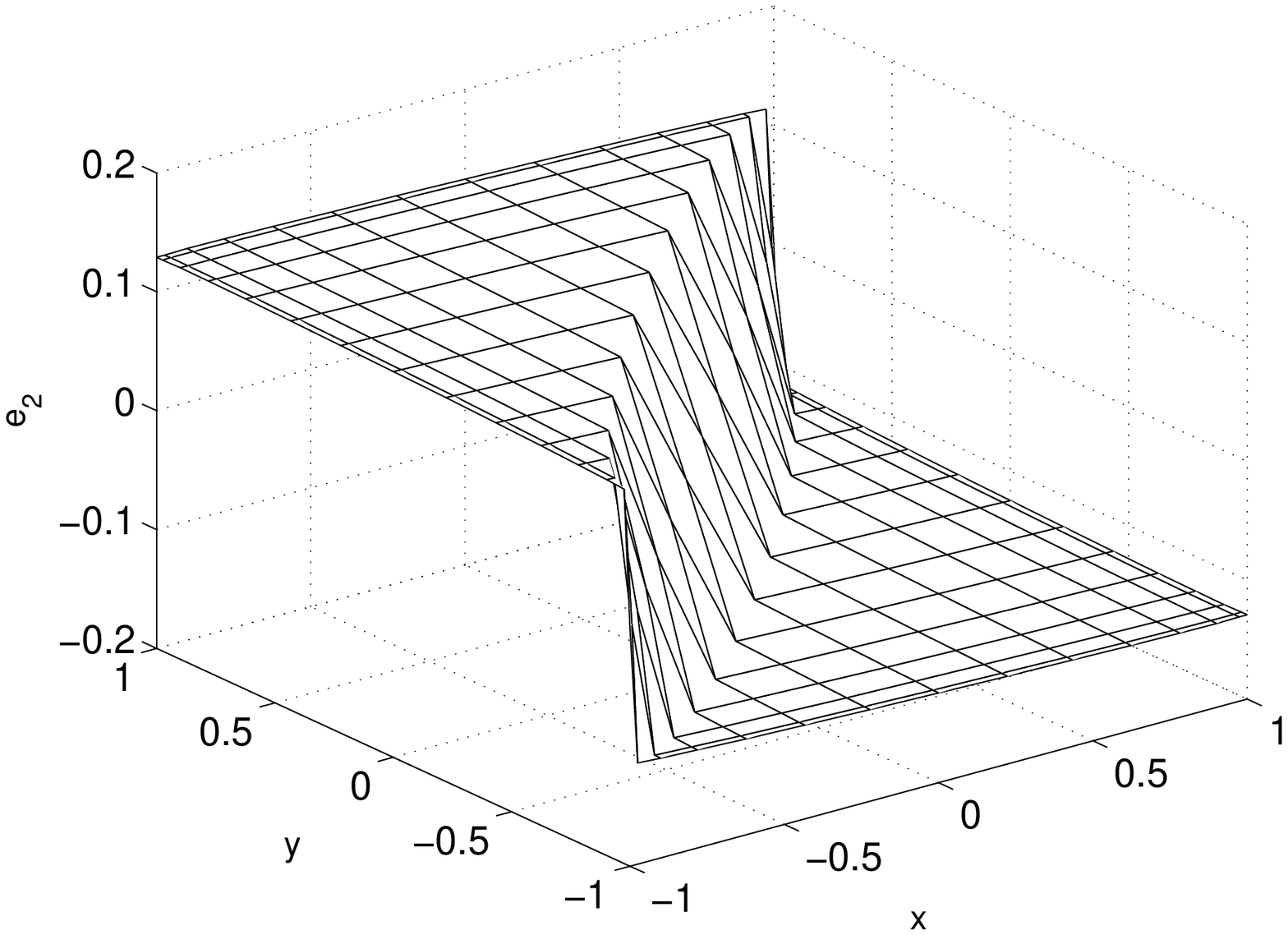}
  \includegraphics[width=7cm, height=7cm]{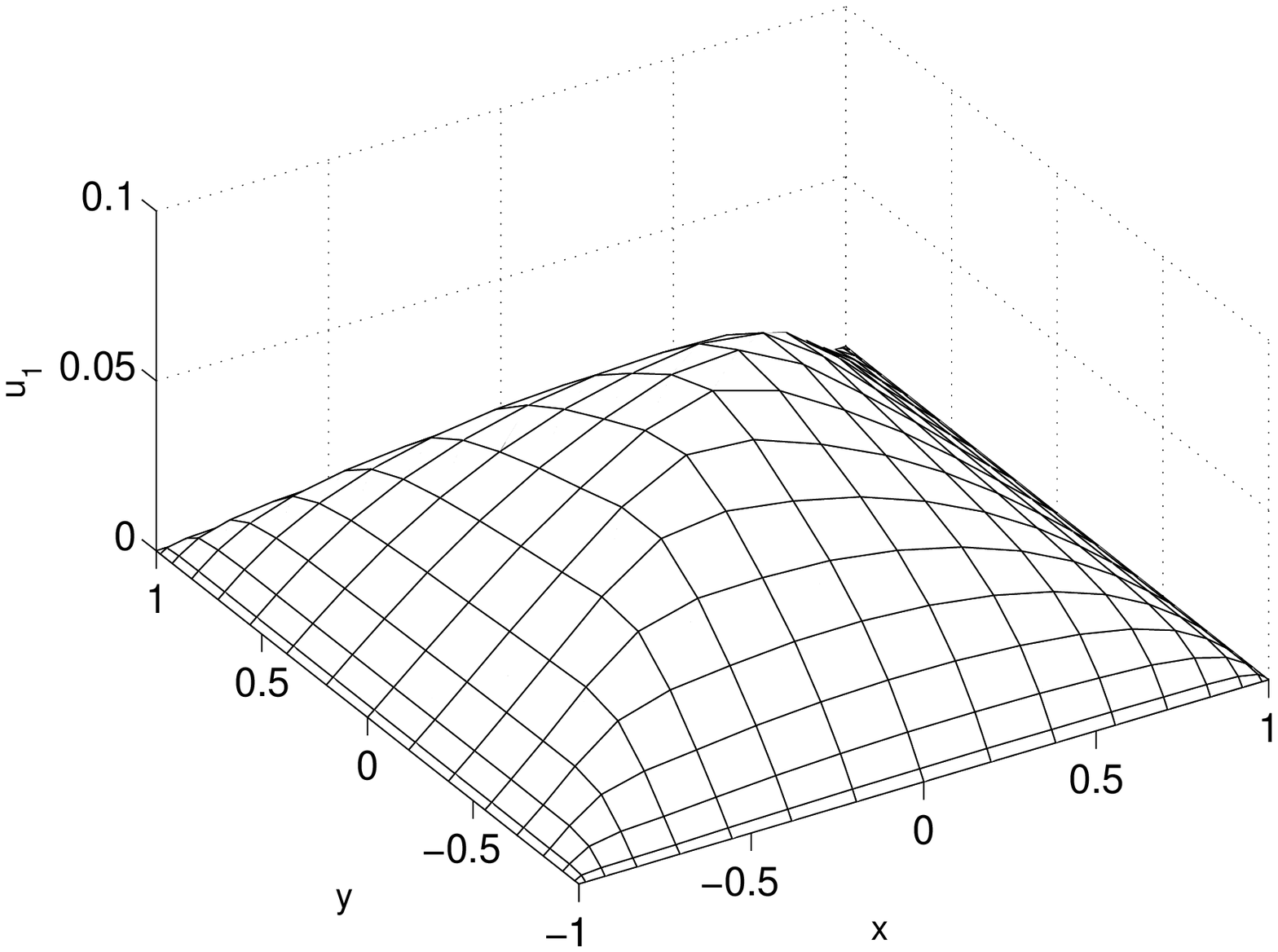}
  \includegraphics[width=7cm, height=7cm]{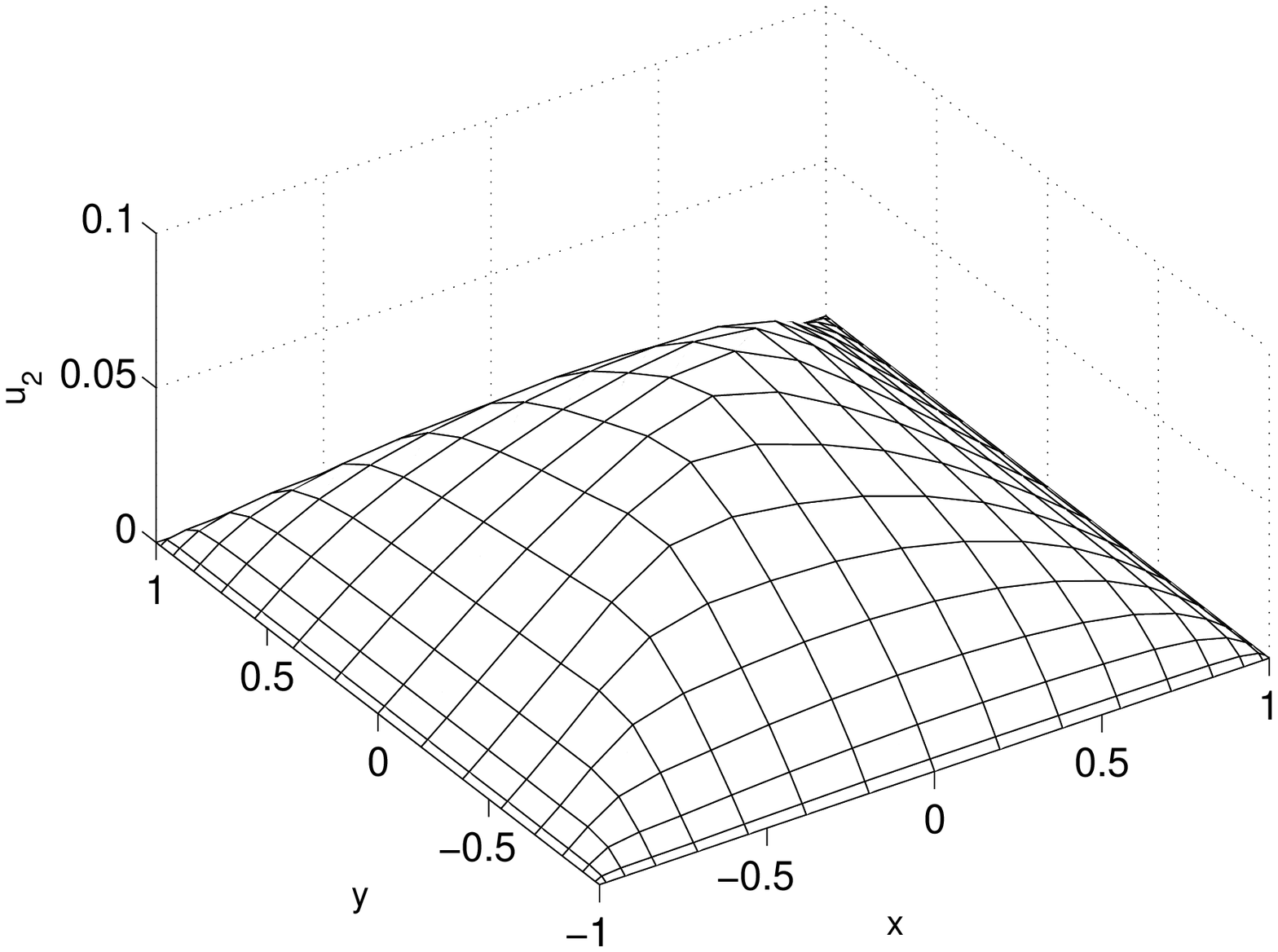}
  \caption{Phase combination in a SMA patch induced by mechanical loading
along the $x$ and $y$ directions.}
    \label{Case2}
  \end{center}        \end{figure}

The final phase combination estimated in this case are presented in
Fig.\ref{Case2}.   The distribution of the order parameter $e_2$ is given in the upper
right subplot.  For comparison purpose, the estimated distribution of $e_2$
from the GA is presented in the upper left subplot.
The final  displacement distributions along the $x$ and $y$ directions are
presented by the two lower subplots.  Once again, the entire domain 
can be divided by two parts with values $e_2 \simeq 0.11$ 
$e_2 \simeq -0.11$,  respectively, referred to as martensite plus and martensite
minus \cite{Melnik2001,Wang2004b}.

\begin{figure}   \begin{center}
  \includegraphics[width=7cm, height=7cm]{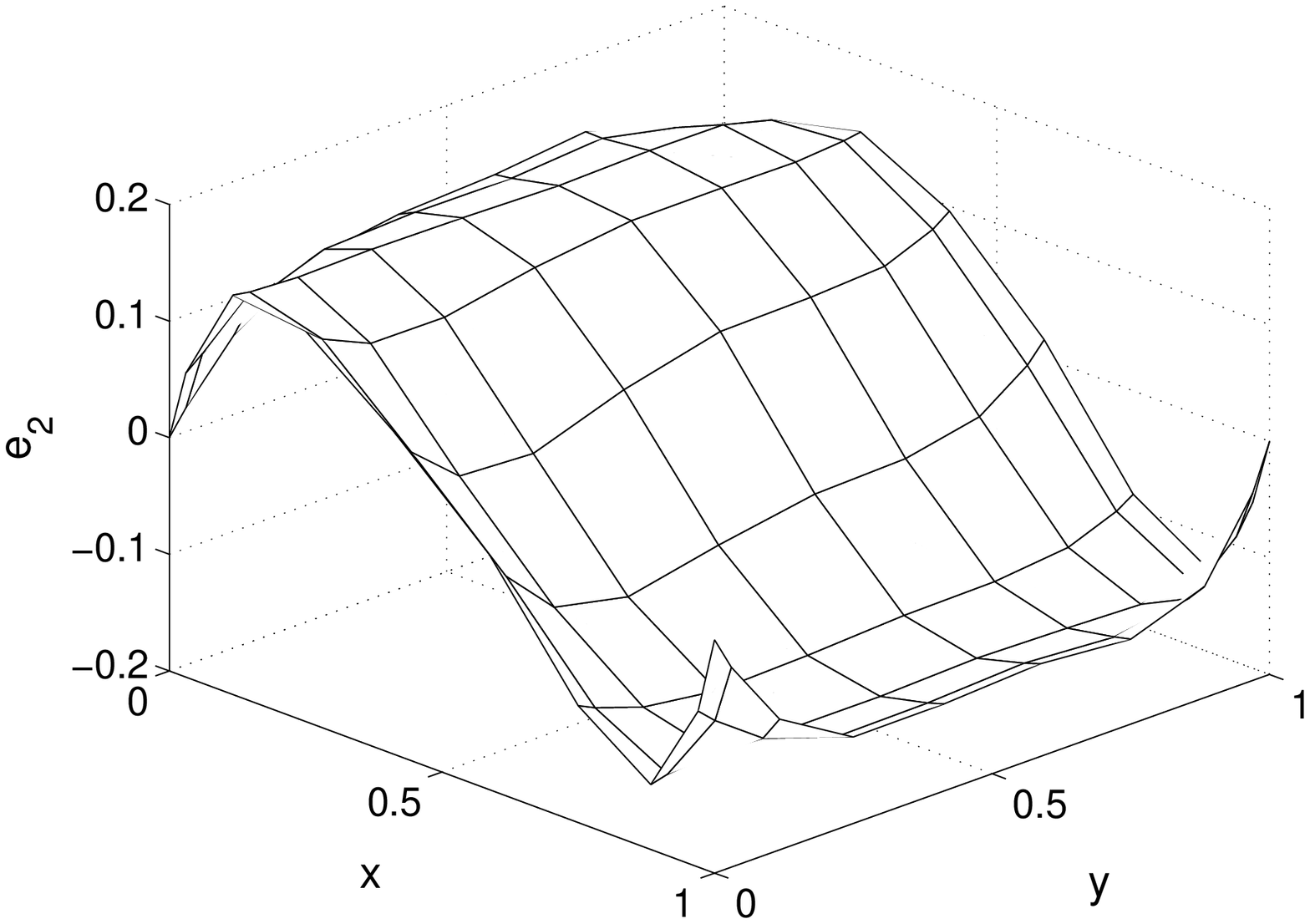}
  \includegraphics[width=7cm, height=7cm]{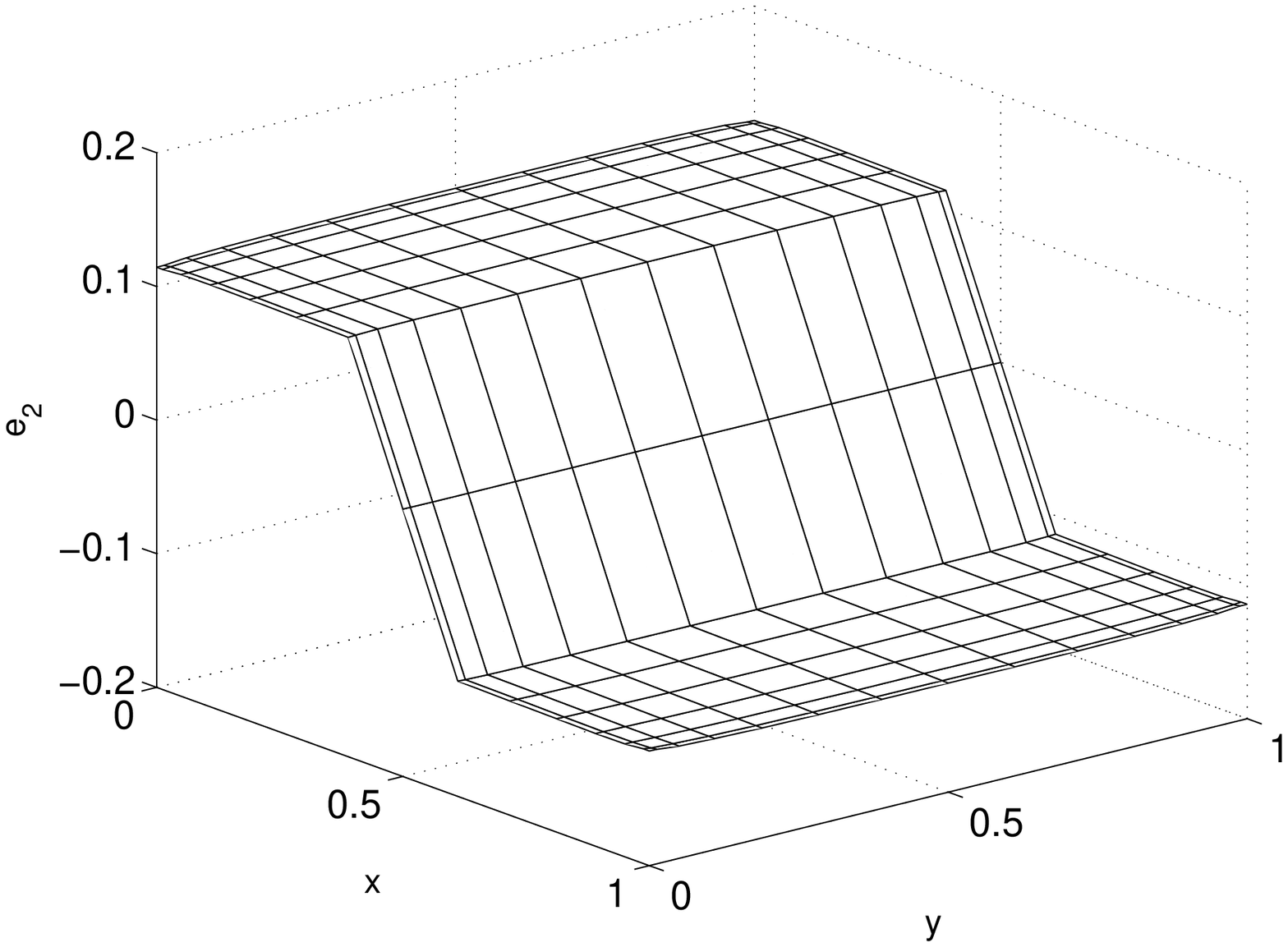}
  \includegraphics[width=7cm, height=7cm]{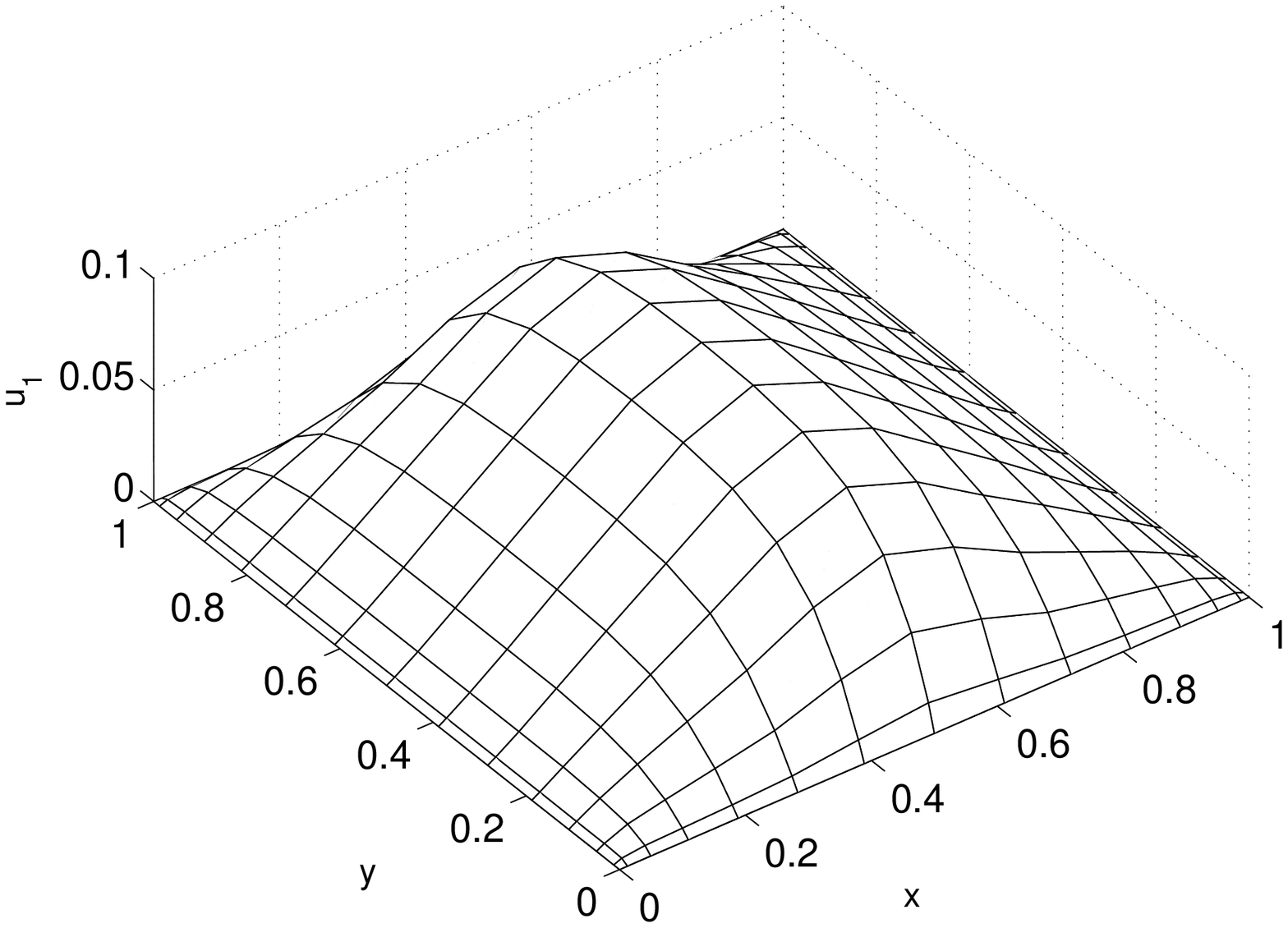}
  \includegraphics[width=7cm, height=7cm]{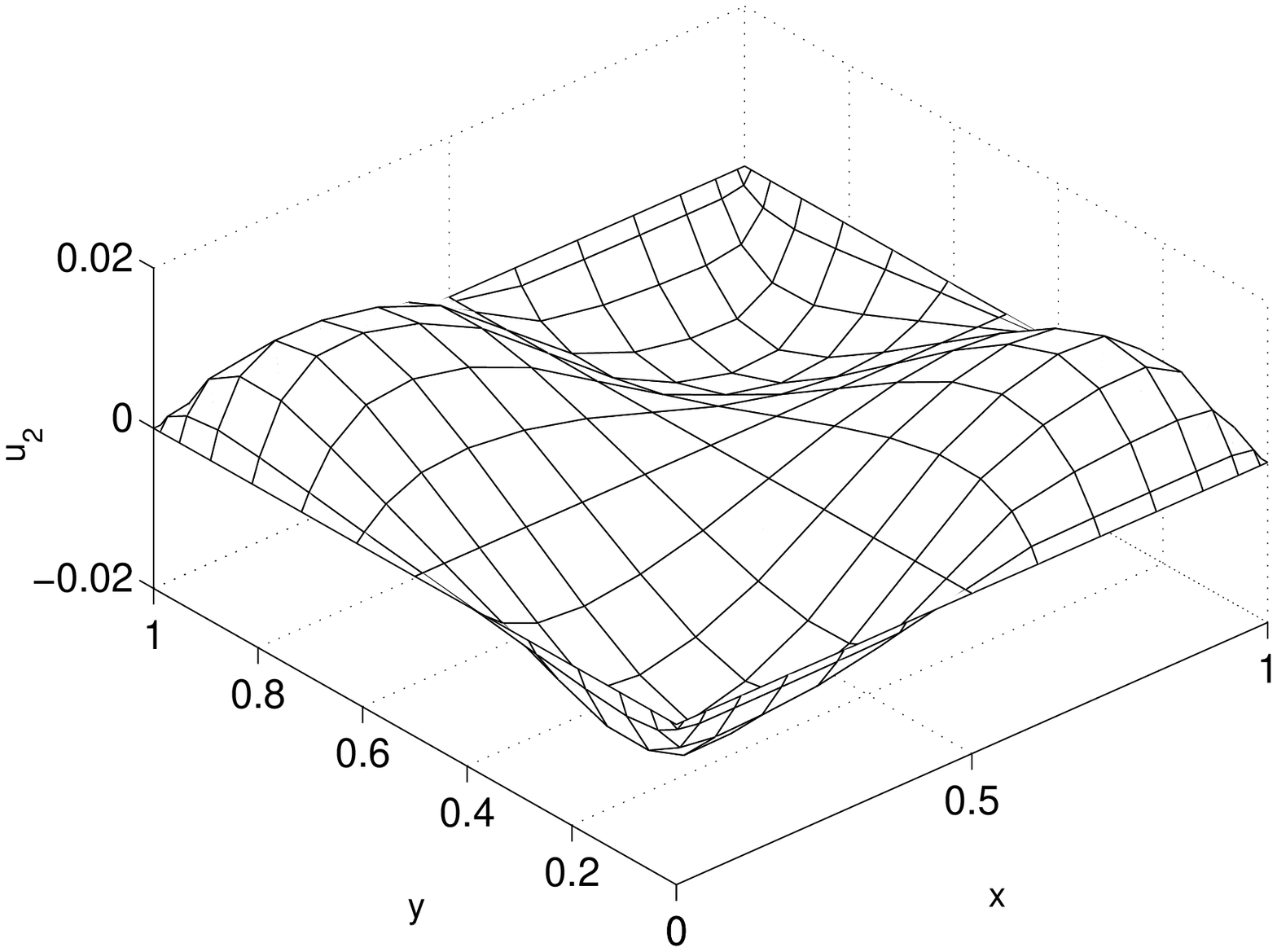}
  \caption{ Phase combination in a SMA patch induced by mechanical loading
along only the $x$ direction}
    \label{Case3}
    \end{center}       \end{figure}

In the last  experiment,  we consider the same patch  $[0,1]\times [0,1] cm^2$
 with modified distributed forces  $f_y=0, f_x= 2000
 g/\left(ms^{2}cm^{2}\right) $ (with the same initial conditions as in the
 previous experiment). All other
computational parameters are taken the same as those in  the previous experiment.
The numerical results are presented in Fig.\ref{Case3} in a  way similar to
 already reported.  In this case, the entire structure is divided
into martensite plus and minus in a different way, the interface is the central vertical
line due to the horizontal symmetric loading. The simulated order parameter is
still close to either $0.11$ or $-0.11$ and the estimated $e_2$ distribution
 obtained with the GA can capture the essence of its final profile.

In the analysis that follows we explain why the order parameter takes values close to either  $0.11$ or
$-0.11$. From the Landau free energy
function, it is easy to estimate the order parameter value which minimizes the
local free energy by setting:
      \begin{equation}      \frac{ \delta F_l }{\delta e_2} = 0.
\end{equation}

 For the 1D case, the order parameter should
be replaced by $\epsilon$.  A simple operation gives the following equation:
      \begin{equation}
        a_2 (\Delta  \theta)e_{2} - a_4e_{2}^3 + a_6e_{2}^5 = 0,
\label{last}
   \end{equation}

 \noindent   which means:
   \begin{equation}
    e_2 =  \frac{ a_4 \pm \sqrt{a_4^2-4a_6a_2 \Delta \theta}}{2a_6}.
   \end{equation}

\noindent Note that $e_2=0$  is also a solution to (\ref{last}) and can be associated with
the austenite phase which is unstable at the current temperature.  
 Therefore, martensitic phases in this case are of greater interest.
The  temperature difference from
the transformation temperature here is given as $\Delta \theta = \theta - \theta_0=2^o$,
and the local minima can be estimated as:  $ e_2 =\pm 0.1146.$

All three numerical experiments have demonstrated that  the
distributions of the order parameter are represented by a combination of two different
values,  $ e_2 = \pm 0.11$, which agrees well with the prediction of the above
analysis.


\section{Conclusions}

In the present paper, phase combinations in the 1D and 2D SMA structures have
been analyzed in the case of square to rectangular transformations.   The
phase combinations  have been obtained by minimizing the
bulk energy in the considered  SMAs structures, subject to given temperature
distribution, mechanical loadings, and boundary conditions. 
By combining the global and local search  techniques, the developed hybrid optimization method
provides a promising strategy for the solution  of the associated non-convex problem.



\newpage

\newpage

  \newpage

  \newpage

  \newpage

\end{document}